\documentclass[preprint,12pt, a4paper]{elsarticle}
\usepackage[utf8]{inputenc}
\usepackage[margin=1.5cm]{geometry}
\usepackage{titlesec}
\usepackage{tabu}
\usepackage{enumitem}
\usepackage{amssymb}
\usepackage{xcolor}
	
\usepackage{soul}

\usepackage{graphicx} 

\newlist{selectlist}{itemize}{2}
\setlist[selectlist]{label=$\square$,leftmargin=*,noitemsep,topsep=0pt}

\usepackage{lmodern}

\usepackage{hyperref}
\hypersetup{
    colorlinks=true,
    linkcolor=blue,
    filecolor=magenta,      
    urlcolor=blue,
}
 
\titleformat{\section}[block]{\hspace{1em}\bfseries}{\thesection.}{0.5em}{} 
\titleformat{\subsection}[block]{\hspace{1em}}{\thesubsection}{0.5em}{}

\begin{document}
\vspace*{0.2in}

\begin{flushleft}
{\Large
\textbf\newline{\textbf{\textit{\textit{HappyFeat}} - An interactive and efficient BCI framework for clinical applications}}
}

\bigskip

Arthur Desbois\textsuperscript{1},
Tristan Venot\textsuperscript{1},
Fabrizio De Vico Fallani\textsuperscript{1},
Marie-Constance Corsi\textsuperscript{1,*}

\bigskip
\textbf{$^1$} Sorbonne Universite, Paris Brain Institute - ICM, CNRS, Inria, Inserm, AP-HP, Hopital Pitie Salpetriere, F-75013, Paris, France
\\
\bigskip

* Corresponding authors: arthur.desbois@gmail.com; marie-constance.corsi@inria.fr

\end{flushleft}
\bigskip

\section*{Abstract}

\begin{small}

Brain-Computer Interface (BCI) systems allow to perform actions by translating brain activity into commands. Such systems require training a classification algorithm to discriminate between mental states, using specific features from the brain signals. This step is crucial and presents specific constraints in clinical contexts. 

\textit{HappyFeat} is an open-source software making BCI experiments easier in such contexts: effortlessly extracting and selecting adequate features for training, in a single GUI. Novel features based on Functional Connectivity can be used, allowing graph-oriented approaches. We describe HappyFeat’s mechanisms, showing its performances in typical use cases, and showcasing how to compare different types of features.

\end{small}

\hspace{1cm}

\hspace{1cm}

\textbf{Keywords} - BCI, Signal Processing, Classification, Motor Imagery, Functional Connectivity, Brain Networks 

\newpage


\section*{Introduction}


Brain-Computer Interface (BCI) systems allow to transcribe brain signals into commands. For this purpose, classification algorithms are used to discriminate between different mental states \cite{lotte_review_2018}. The field of possible applications is vast, ranging from communication to prosthesis control and post-stroke rehabilitation\cite{pfurtscheller_future_2006}. Multiple BCI paradigms exist, such as P300 and Steady-State Visual Evoked Potentials (SSVEP)\cite{lotte_electroencephalography_2015}. We choose to focus on Motor Imagery (MI), as it is prominent in prospective therapeutic usages (e.g. post-stroke rehabilitation\cite{pichiorri_brain-computer_2017}), which aligns with our current research efforts aiming to improve the use BCI in clinical settings. In MI, the subject actively imagines a movement without actually performing it, in order to command a virtual or real device (e.g. moving an object on a screen, controlling a robotic arm). Consequently, MI offers a high level of interactivity and requires a strong active implication from the subject.

\begin{figure}[ht!]
	\centering
	\includegraphics[width=0.65\textwidth]{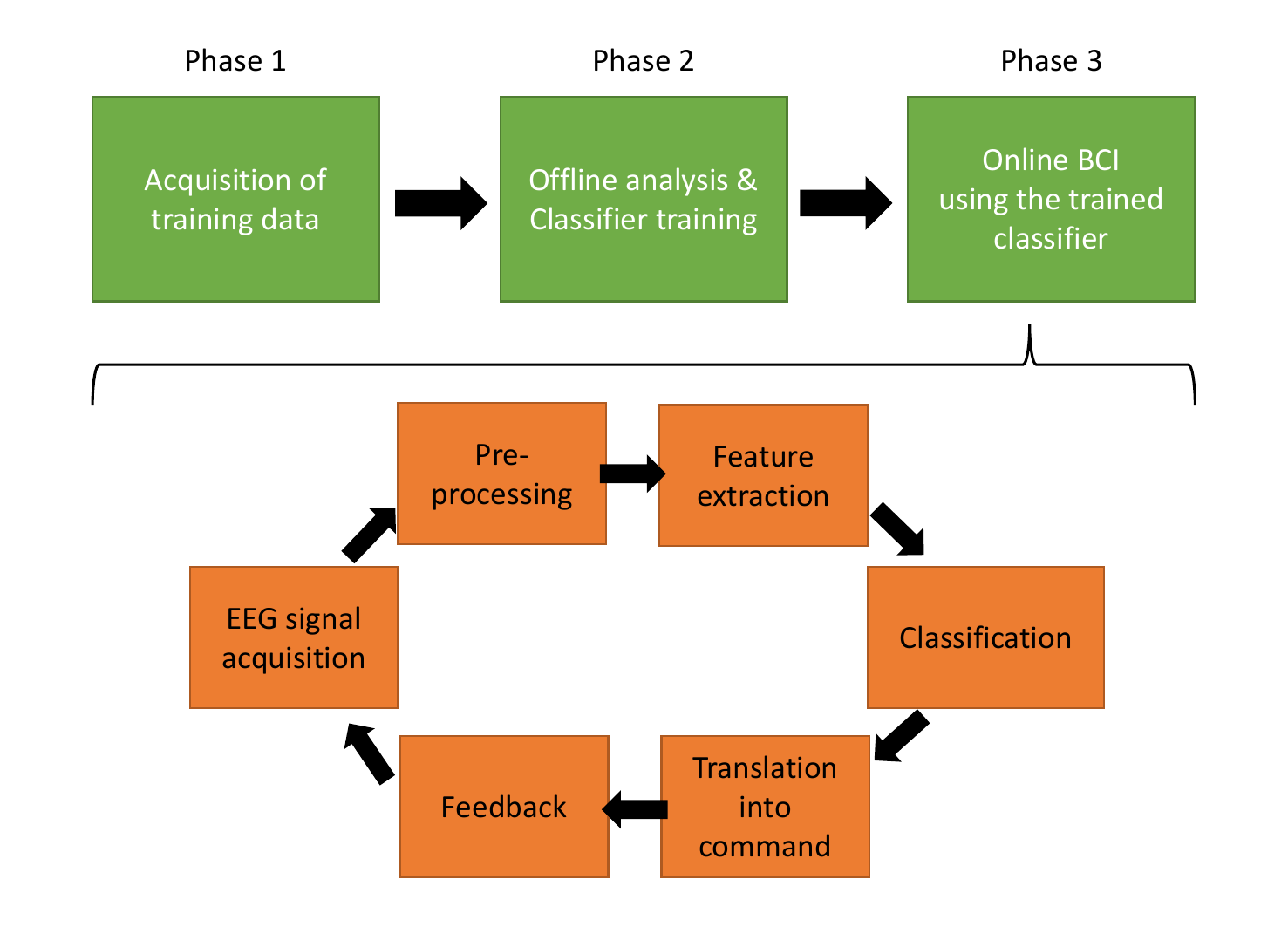}
	\caption{\textbf{Typical BCI protocol}.}
	\label{fig:bci}
\end{figure}

MI protocols consists of multiple phases\cite{pfurtscheller_motor_2001}\cite{jeunet_why_2016} (illustrated in \textit{Figure \ref{fig:bci}}): an\textit{ acquisition phase} of training data; an \textit{offline analysis phase} usually including pre-processing, extraction of features of interest (based on e.g. power spectra, functional connectivity), feature selection and classification algorithm training; a closed-loop \textit{online BCI} usage using the trained classification algorithm.


The performance of a BCI system, while dependent on internal (e.g. concentration, fatigue\cite{myrden_effects_2015} and ease with BCI) and external factors (e.g. montage of electrodes), is strongly linked to the correct training of the classification algorithm. Therefore, the choice of adequate features that capture the user's intent is crucial.

 
The offline analysis phase, leading to the choice of those features, should be made as short and efficient as possible for two reasons:
\begin{itemize}
	\item A complete MI experiment is a long and strenuous process lasting multiple hours, along which the subject's concentration and motivation can drop. Patients' conditions increase the need for a fast feature selection process. 
	\item Too long a time between the different phases may result in low classification accuracy, and therefore poor BCI performance. Indeed, as time passes, many physical parameters may change: the impedance or localization of Electroencephalographic (EEG) sensors (leading to change in EEG signals characteristics and quality), the subject's motivation, but also their mental activity itself. These changes mean that at the time of the final online classification phase, the subject's EEG signals and brain activity might not match with the ones used by the BCI experimenter to select features and train the classifier anymore.
\end{itemize} 

Therefore, there is a crucial need for providing assistance to the experimenter during the training phase, to identify the most relevant and robust features.

Software solutions already exist today to manipulate BCI systems, or to analyze acquired EEG data. For creating and manipulating complete BCI systems, one can cite OpenViBE\cite{renard_openvibe_2010}, which offers compatibility with a wide range of EEG hardware, practical modular tools to edit BCI systems as "scenarios", and powerful signal processing capabilities. It also allows interfacing with Virtual or Augmented Reality headsets and other virtual environments such as video games. Timeflux\cite{clisson_timeflux_2019} is another software solution, allowing experimenters to acquire and process EEG signals in real-time, with a high degree of scenario and interface flexibility. Another existing solution is BCI2000\cite{schalk_bci2000_2004}, which allows a lower level of customization and modularity. However, none of the aforementioned solutions provide tools to interactively identify the best features to use for training and fine-tuning a BCI system.

Tools for offline analysis of EEG signals also exist today, such as Brainstorm \cite{tadel_brainstorm_2011}, EegLab\cite{delorme_eeglab_2004} or FieldTrip\cite{oostenveld_fieldtrip_2010}. While these software embark wide selections of statistical and signal processing methods and are useful for research on neurophysiological phenomena, they lack capabilities in the domain of machine-learning, and the possibility to interface with BCI in a direct workflow.

Finally, the open-source software package MNE-Python\cite{gramfort_mne_2014} provides extensive functionalities allowing experimenters to manipulate EEG data and analyze them with signal-processing and machine-learning methods using scikit-learn\cite{pedregosa_scikit-learn_2011}, some of which  \textbf{\textit{HappyFeat}} makes use of.


\section*{Description and features}

\textbf{\textit{HappyFeat}} is a Python framework, consisting in a main application and various scripts and automation routines, allowing to facilitate the use of MI-based BCI pipelines. This is achieved by grouping all necessary manipulations and parameters in an unified graphical user interface (GUI), and making the steps of feature extraction, feature selection and classifier training as effortless and seamless as possible, so that experimenters may focus on building an efficient BCI.

The following contains a description of \textit{HappyFeat}'s main features and mechanisms.

\subsection*{\underline{Key features}}

\begin{figure}[ht!]
	\begin{minipage}{\textwidth}
		\centering
		\includegraphics[width=\textwidth]{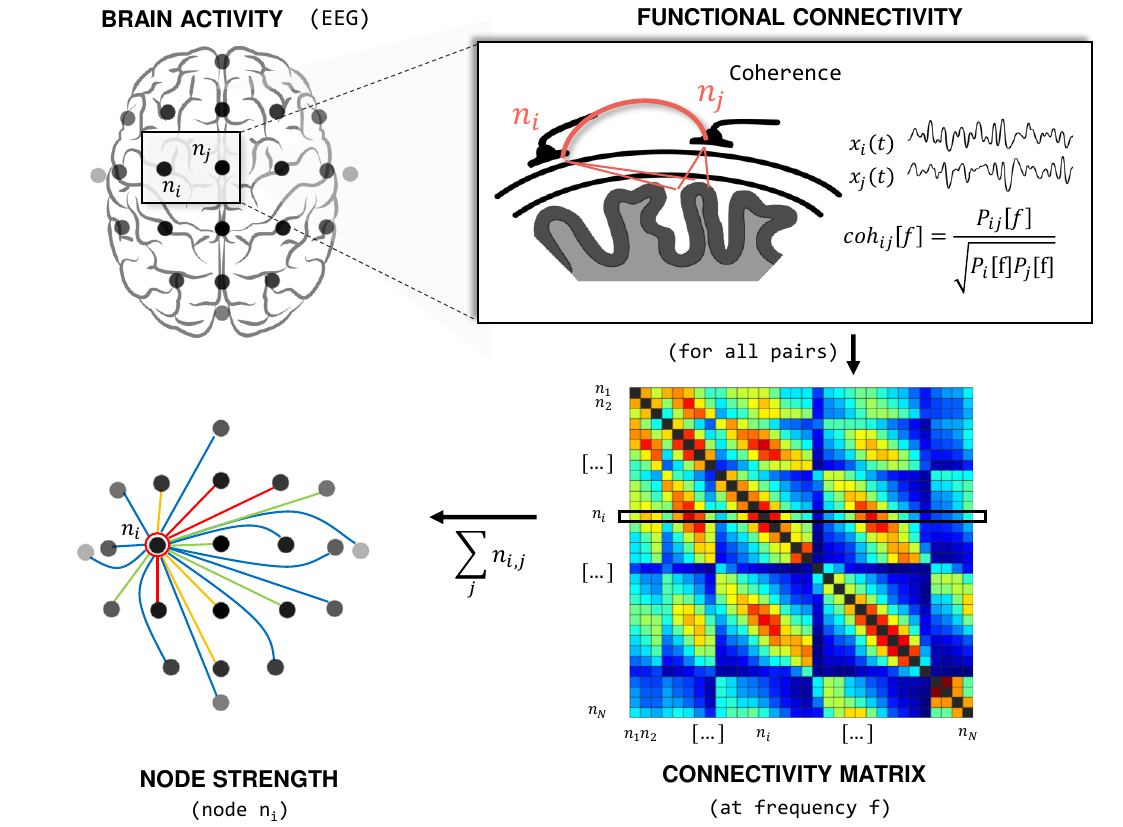}
		\caption{\textbf{From brain activity to Functional Connectivity and Networks}. Spectral coherence (or imaginary coherence) between EEG signals from electrodes $n_i$ and $n_j$ is computed from their Power Spectral Densities and Cross Spectral Density. Applying this operation to every electrode pairs yields a connectivity matrix, visualized here at frequency $f$. From this matrix, network estimators can be extracted. The "Node strength" of a given node is obtained by summing all connections to this node. Computing the node strength for all frequencies in the considered range yields a "Node Strength Density" which can be analyzed and manipulated similarly to the usual Power Spectral Density. Courtesy of J. Gonzalez-Astudillo\protect\cite{gonzalez_astudillo_development_2022} }
		\label{fig:nodestrength}
	\end{minipage}
\end{figure}

\begin{itemize}	
	\item \textit{HappyFeat} is a \textbf{software assistant for feature extraction and selection}. It proposes an exploration-oriented workflow, where experimenters can extract, visualize and select Features of Interest (FOI) for training as many times as needed, in a short time, until a satisfying classification training accuracy is reached.
	
	\item Every operation from signal loading and feature extraction to classifier training is handled from a \textbf{unified, dashboard-like GUI}, removing the need to use different softwares for data acquisition, feature analysis, classifier training and online classification, and to manage data formatting across the different environments.
	
	\item Along with the commonly used \textbf{Power Spectral Density} (PSD), \textit{HappyFeat} enables to work with \textbf{Functional Connectivity}, allowing to use \textbf{novel network-based approaches} based on recent research\cite{gonzalez-astudillo_network-based_2021}\cite{cattai_characterization_2018}. This approach is illustrated in \textit{Figure \ref{fig:nodestrength}})
	
	\item \textit{HappyFeat} uses \textbf{OpenViBE} in the background for the extraction and training parts, as a fast and efficient processing engine, taking advantage of its optimized C++ implementation of signal processing methods (notably using the Eigen\footnote{\url{https://eigen.tuxfamily.org/index.php?title=Main\_Page}} library). The generation and manipulation of OpenViBE scenarios is entirely \textbf{automated} via scripts and templates, removing the inherent risk of mistakes in a time constrained environment. Feature visualization and selection use tools from MNE-Python\cite{gramfort_mne_2014} and scipy\cite{virtanen_scipy_2020}. The GUIs are built with PyQt\footnote{\url{https://doc.qt.io/qtforpython-5/}}.
	
	\item \textit{HappyFeat} puts the emphasis on \textbf{reproducibility}, by keeping track of all manipulations (EEG sessions file lists, signal processing steps and parameters, classification attempts) and allowing to save, load and export previous work.
\end{itemize}

Two main use-cases are targeted:
\begin{itemize}
	\item Using MI in a clinical setting (e.g. stroke rehabilitation), by greatly reducing the risks of mistakes during the offline analysis and the time needed to perform this step, quickly bridging the gap between EEG data acquisition and online BCI usage. 
	\item Exploring new, alternative metrics for discriminating between mental states. To this aim, prototypes for prospective methods need to be validated on signal databases, before moving on to experimental conditions. \textit{HappyFeat} helps bridging this gap, and provides a framework in which such methods can be tested, after implementation.
\end{itemize}

\subsection*{\underline{Mechanisms}}

\begin{figure}[ht!]
	\centering
  	\makebox[\textwidth][c]{\includegraphics[width=\textwidth]{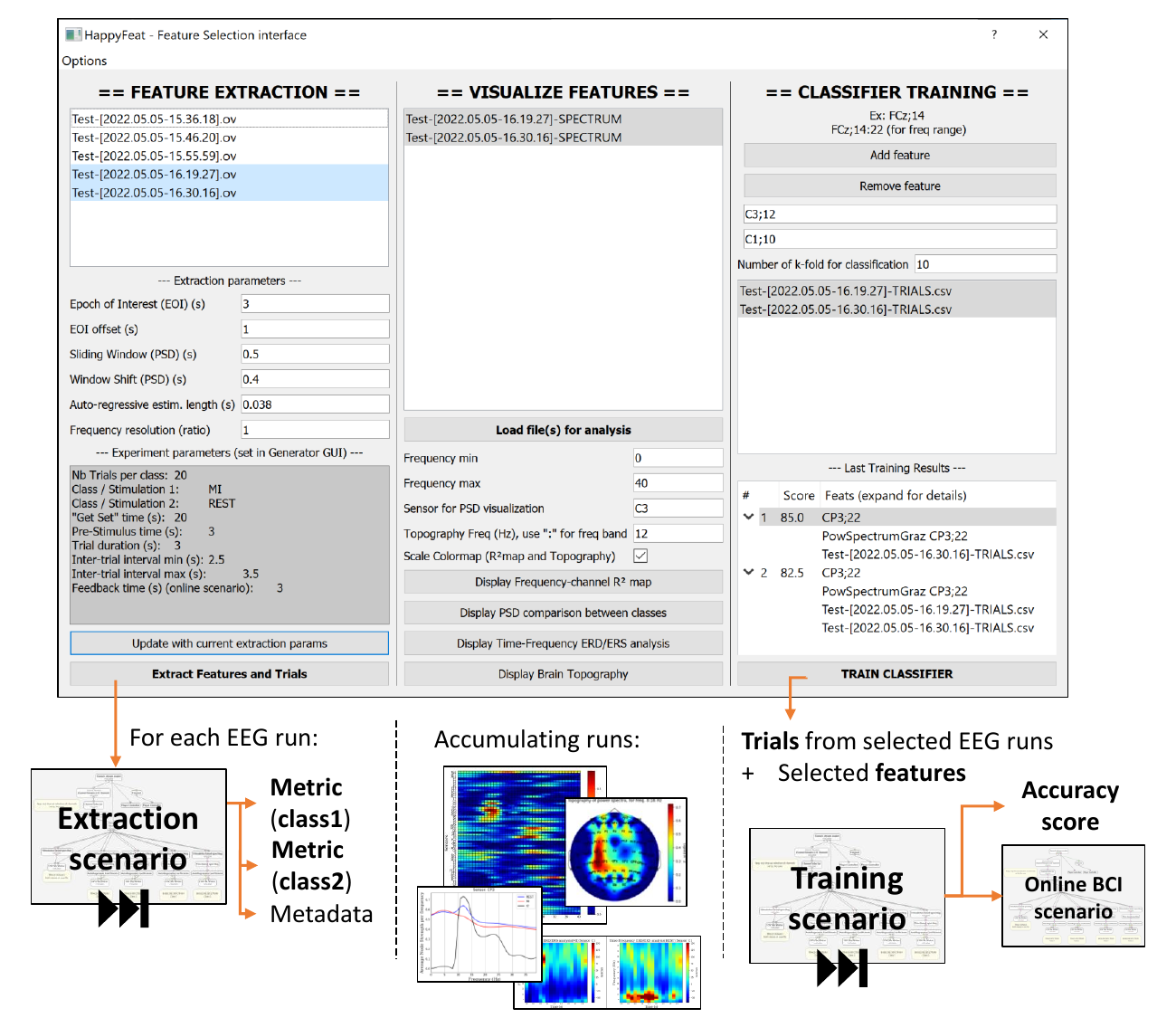}}
  	
	\caption{\textbf{HappyFeat main GUI}. The interface is split in 3 parts: in the leftmost part, experimenters can perform \textit{Feature Extraction} from recorded EEG signals available in the workspace. The central part is dedicated to \textit{Feature Visualization} for signals that have undergone extraction, allowing to select adequate features for training. The last and rightmost part allows the experimenter to train the classifier using selected features and signal files. Along the way, certain parameters can be set, while other parts of the interface help to remind acquisition settings or last training attempts.}
	\label{fig:hf1}
\end{figure}

\begin{figure}[ht!]
	\centering
	\includegraphics[width=\textwidth]{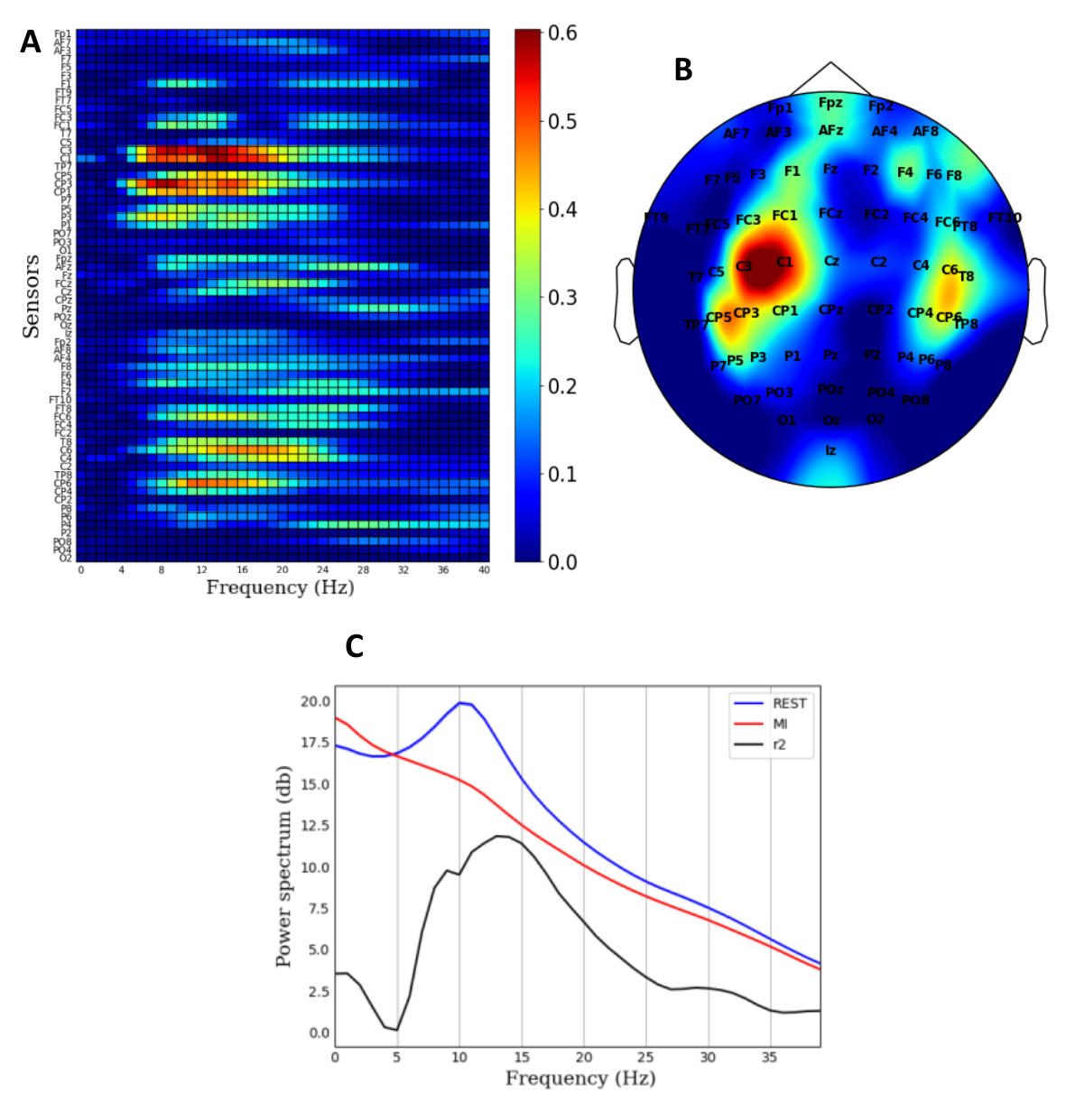}
	\caption{\textbf{Comparison of MI conditions "Rest" and "MI" in Subject 1, using PSD}. $R^2$ values represented as a frequency-sensor map (A), as a ”brain topography” mapped on a scalp (at  frequency 12Hz) (B), and direct comparison of PSD averaged over trials (at sensor C1) (C). The PSD is obtained using Burg's AR method. Figures A and B share the same color scale for $R^2$ values, and show desynchronization in C1, C3, CP3, CP5 sensors, in the alpha and low beta bands (from roughly 8Hz to 18HZ). Fig. C highlights this desynchronization with more precision in the 11 to 15Hz band. The black curve in C shows the $R^2$ values, scaled from 0 to 1.}
	\label{fig:hf_fig_sub1_psd}
\end{figure}

\begin{figure}[ht!]
	\centering
	\makebox[\textwidth][c]{\includegraphics[width=\textwidth]{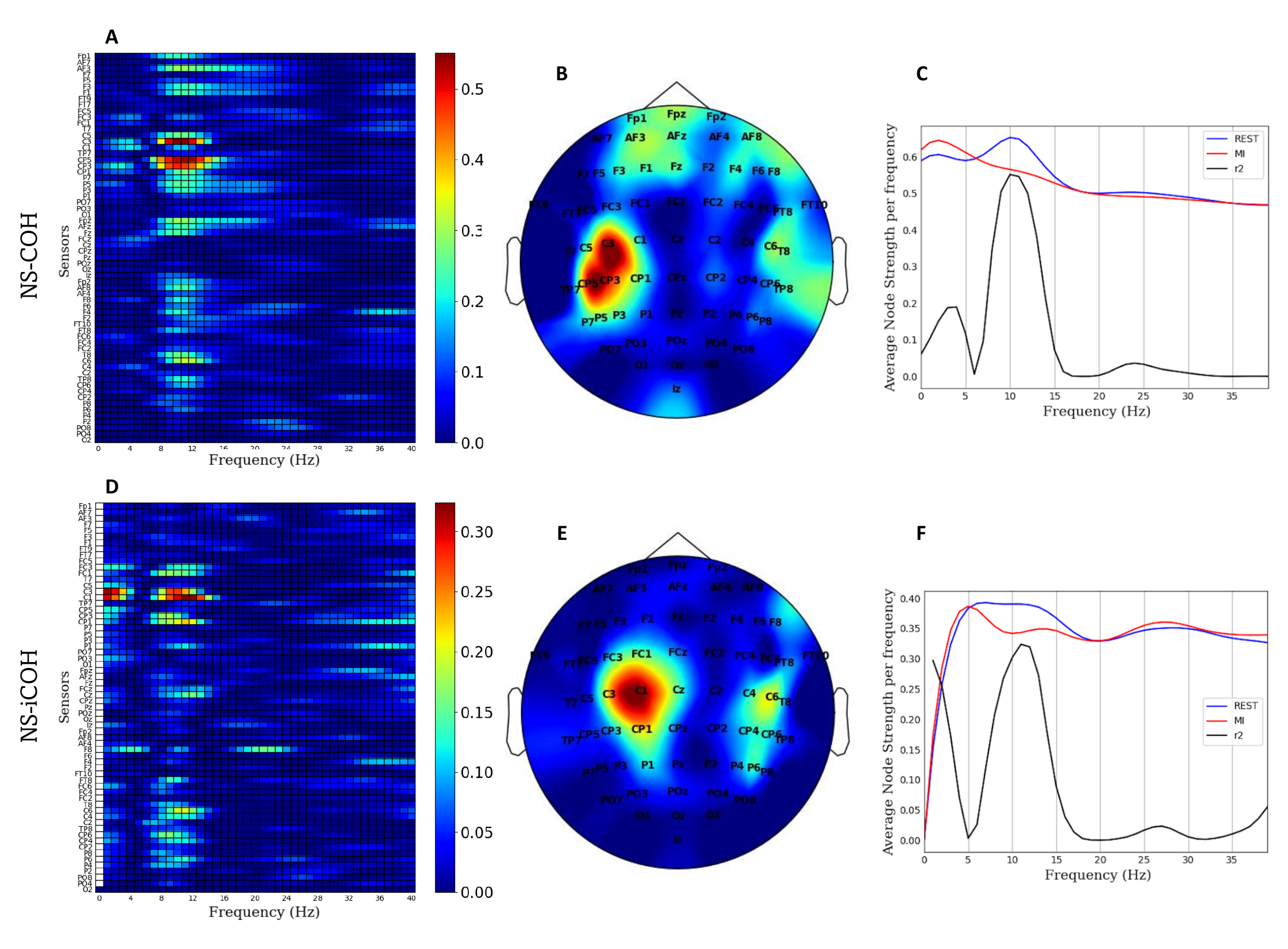}}
	
	\caption{\textbf{Comparison of MI conditions "Rest" and "MI" in Subject 1, using NS-COH (top row) and NS-iCOH (bottom row)}. $R^2$ values represented as a frequency-sensor map (A, D), as a ”brain topography” mapped on a scalp (B, E) at frequency 12Hz, and direct comparison of NS averaged over trials for sensor C3 (C) and C1 (F). Figures A and B share the same color scale for $R^2$ values, as do figures D and E. The black curves in C and F show the $R^2$ values, scaled from 0 to 1. }
	\label{fig:hf_fig_sub1_coh}
\end{figure}

\textit{HappyFeat}'s main GUI and mechanisms are shown in Figure \ref{fig:hf1}
\begin{itemize}
	\item \textit{HappyFeat} allows the experimenter to choose between different metrics for discriminating between mental states, such as Power Spectral Density (PSD)\cite{diez_comparative_2008} or network estimators based on Functional Connectivity (e.g. Node Strength)\cite{nolte_identifying_2004, cattai_characterization_2018, cattai_phaseamplitude_2021}, both of which are described in the Annex section\ref{sec:annex}. Experimenters can also use a mix of two different estimators, allowing to classify using both PSD and Node-strength features for example.
	
	\item \textit{Data loading}: Experimenters can select different EEG recordings (either directly after acquisition during a BCI experiment, or using pre-recorded signals) from which to compute metrics and extract features. Parameters relevant to the chosen classification metric can be edited. OpenViBE scenarios are automatically created, updated and ran in the background using relevant information provided by the experimenter, without needing any additional manipulation.
	
	\item \textit{Visualization tools} allow to analyze and select features of interest, accumulating statistics across selected EEG runs. Such figures allow for a comfortable and easy selection of adequate features for training the classification algorithm. Experimenters may open any number of visualization windows, allowing to compare $R^2$ values between MI conditions (e.g. \textit{MI} vs \textit{REST}) as a channel-frequency map, as power densities for a given channel, as a topography map for a given frequency range, etc. Examples of such visualizations are given in Figures \ref{fig:hf_fig_sub1_psd} and \ref{fig:hf_fig_sub1_coh}.
	
	\item \textit{Classifier training} can be done iteratively in a trial-and-error way. Using a set of EEG runs (from which features have been extracted previously in \textit{HappyFeat}) selected by the experimenter, and FOIs selected in the previous step, a classification algorithm (such as Linear Discriminant Analysis (LDA)\cite{sanei_classification_2013}) is trained in a few seconds, and the application provides a training accuracy score. 

	At this point, if the training accuracy score is satisfactory, experimenters can proceed with the last step of the experiment (i.e. the closed-loop online BCI), using an OpenViBE scenario automatically updated with the trained classifier and the selected features. In the case of an insufficient score, experimenters may simply go back to the previous steps, either directly trying other features to try training the classifier again; or modify their visualizations to select other features; or extract features from other EEG signals. Going back and forth between these steps only takes a few seconds, and manipulations are limited to the strict minimum.
		
	\item A \textit{Session \& Settings Management System} allows to export and import \textit{"workspaces"}, allowing experimenters to keep track of manipulations previously done on EEG recording sessions: extraction results and corresponding parameters, training attempts and accuracy scores with the corresponding feature set, etc. 
			
\end{itemize}

\subsection*{\underline{Example of metric comparison}}

The common metric used in MI protocols to discriminate between mental tasks is the Power Spectral Density (PSD), which is available in \textit{HappyFeat}. Alternative metrics based on functional connectivity\cite{cattai_characterization_2018, cattai_phaseamplitude_2021} are proposed. As the raw connectivity matrices are difficult to interpret to physically store, having a dimension of \textit{(nb.channels $\times$ nb.channels $\times$ nb.frequencies)}, the metrics proposed in \textit{HappyFeat} are network metrics, such as the \textit{node-strength}, obtained by summing the weights of all connections for each channel. The resulting matrix of node-strengths has a dimension of $(nb.channels \times nb.frequencies)$, which can be analyzed using the same tools as the PSD. Similarly to the PSD, sets of (channel, frequency) can be used as features for training the classification algorithm. \textit{HappyFeat} also proposes to use a mix of different metrics (i.e. PSD and Node-Strength) to train the classifier.

To illustrate this, two subjects from Venot et al.'s BRACCIO protocol\cite{venot_mental_2023} were selected (both male, aged 25, right-handed). In this protocol, subjects were asked to perform either MI of the right hand, or no MI ("Rest"), with visual feedback in the form of a moving robotic arm. 120 trials (60 per class) were used.

Table \ref{tab:table_acc} gives the accuracies obtained for both subjects when training a classifier with particular features, using PSD, coherence-based node strength (NS-COH), imaginary part of coherence -based node strength (NS-iCOH) and mixing metrics.

\textit{Figures \ref{fig:hf_fig_sub1_psd} and \ref{fig:hf_fig_sub1_coh}} show the channel-frequency $R^2$ maps and metric comparison for Subject 1, obtained with \textit{HappyFeat}'s visualization tools.

A high inter-subject variability can be observed. Subject 1 shows similar performances between metrics, with a slight advantage for PSD. Subject 2 illustrates how using NS-COH, NS-iCOH and mixing them with PSD can lead to performance improvement.

\begin{table}[h!]
	\begin{center}
		\begin{tabular}{|c|c|c|}
			\hline
			Metric & Feature (channel;frequency) & Training Accuracy (\%) \\
			\hline
			\textbf{Subject 1}&	& \\
			PSD 			&   C3;13			& 91.67	\\
			NS-COH			&  	C3;10			& 81.67	\\
			PSD + NS-COH	&  	C3;13 + C3;10	& 86.67	\\
			NS-iCOH			& 	C1;11			& 80.00  \\
			PSD + NS-iCOH	& 	C3;13 + C1;11	& 91.65  \\
			\hline
			\textbf{Subject 2}&	&  \\
			PSD 			&   C3;11			& 83.35	\\
			NS-COH			&  	C1;12			& 96.65	\\
			PSD + NS-COH	&  	C3;11 + C1;12	& 93.35	\\
			NS-iCOH			& 	P5;11			& 80.00  \\
			PSD + NS-iCOH	& 	C3;11 + P5;11	& 88.35  \\
			\hline
		\end{tabular}		
		\caption{\textbf{Comparison of training accuracies when using connectivity-based features}. }
		\label{tab:table_acc}
	\end{center}
\end{table}

\section*{Impacts}

\textit{HappyFeat} offers a novel complete integrated workflow, allowing to perform all steps of the offline analysis leading to feature selection in BCI setup. Two main use-cases are targeted, whose impacts are methodological and scientific on one side, and clinical and therapeutic on the other.

\begin{itemize}

    \item On the one hand, \textit{HappyFeat}'s potential impact on research using BCI is noteworthy. Novel and innovative algorithms (e.g. based on graph-theory and functional connectivity\cite{cattai_characterization_2018, cattai_phaseamplitude_2021}) can be validated on pre-recorded data, and compared to one another or to state of the art techniques (PSD). \textit{HappyFeat} provides an efficient framework in which new methods for discriminating between mental states can be tested and benchmarked. 

    \item On the other hand, \textit{HappyFeat} helps using MI in a clinical setting (e.g. stroke rehabilitation), by greatly reducing the risks of mistakes during the offline analysis and the time needed to perform this step, quickly bridging the gap between EEG data acquisition and online BCI usage. \textit{HappyFeat}'s mechanisms and graphical interface have been designed to be easily used by experimenters and clinicians without strong programming skills, in order to facilitate the introduction of BCI methods in healthcare. By making therapeutic protocols using BCI feasible and realistically applicable in real-life, \textit{HappyFeat} aims to facilitate the design of innovative training programs to improve neuro-rehabilitation, in order to improve patients' quality of life in the long run.

\end{itemize}

\textit{HappyFeat} has been instrumental in a real-life experimental study using BCI to control a robotic arm\cite{venot_exploring_2022}\cite{venot_mental_2023}, making the acquisition, analysis and BCI steps feasible in the same session, and allowing in a further study on recorded signals to compare training performance of PSD and Spectral Coherence. Such a comparison is illustrated in \textit{figure \ref{fig:hf_fig_sub1_psd}, figure \ref{fig:hf_fig_sub1_coh} and table \ref{tab:table_acc}} with two subjects from this study.

As another example of real-life impact, in the context of the BCINET research project\cite{erc_bcinet_2020}, \textit{HappyFeat} will serve as the cornerstone software for feature analysis with Functional Connectivity. This protocol aims to evaluate training effects and recovery in stroke patients, using BCI and non-invasive stimulation techniques.

\textit{HappyFeat}'s workspace management system allows for reproducible research, by enabling to import or share work sessions between different users. The flexibility offered by metric comparison and the open-source nature of the project will help to develop the use of BCI in multiple research domains and applications not limited to healthcare, and as a pedagogic tools for users new to BCI.


\section*{Limitations \& Perspectives}

\subsection*{\underline{Flexibility}}

In order to offer a safe and risk-free workflow, we chose to build \textit{HappyFeat} around the concept of fixed pipelines, trading OpenViBE's high level of flexibility in designing BCI systems for fixed, efficient pipelines with reproducible results. Nevertheless, even though HappyFeat is designed as a turnkey software solution, more experienced experimenters are free to modify scenarios run along the pipeline and templates used to generate them if the need arises, for example to fine-tune the signal processing chain in the feature extraction step, or to edit the type of feedback provided in the online BCI scenario. An in-depth guide helping to edit \textit{HappyFeat}'s pipelines is available in the software documentation.

\subsection*{\underline{Proposed algorithms and methods}}

At the time of writing, the only machine-learning algorithm proposed for the classification step is \textit{Linear Discriminant Analysis} (LDA). Other methods shall be made available in the future, such as \textit{Support Vector Machine} (SVM) or \textit{Riemannian Geometry} based methods\cite{lotte_review_2018}.

Regarding discriminant metrics based on network and graph theory, the only method available is the node strength, calculated from the coherence-based connectivity matrix. Other network-based metrics will soon be available, such as network laterality or betweenness centrality\cite{gonzalez-astudillo_network-based_2021}. Moreover, \textit{HappyFeat} proposes to mix different metrics, but at the time of writing only PSD + NS-COH or PSD + NS-iCOH may be chosen. A mechanism allowing to mix NS-COH + NS-iCOH or any combination of future implemented metrics will be proposed in a later version of the software.

In practice, feature selection may be realized manually or using automatic method such as Common Spatial Pattern (CSP)\cite{sanei_classification_2013}. However, while such methods offer simplicity and speed, they imply reducing control over the selection process, and a reduced level of interpretability, both of which are crucial when testing prospective feature such as connectivity-based metrics. The workflow proposed in HappyFeat is therefore a trade-off between speed, ease-of-use and human interpretability. It should be noted that such automatic methods could still be proposed in the interface in the future.

A time-frequency ERD/ERS analysis tool is provided in the visualization part of the GUI, but only allows to compare the averaged spectra of trial against a "baseline" defined as the EEG signal acquired right before the stimulation cue (i.e., MI trial vs. MI baseline, and REST trial vs. REST baseline). This tool will be improved in a future version to allow comparing conditions between themselves (i.e. REST vs. MI), and to be based on Morlet wavelets\cite{brodu_comparative_2011}.

\subsection*{\underline{Processing engine, BCI software dependency}}

At the time of writing, \textit{HappyFeat} is built upon the manipulation of OpenViBE scenarios, taking advantage of this software's high level of modularity and its signal processing capabilities. However, work is ongoing to show that other modular BCI software can be used as processing engines (such as Timeflux\cite{clisson_timeflux_2019}), without modifying the mechanisms of HappyFeat.

\section*{Conclusions}

We propose a software to facilitate usage of MI-based BCI for multiple types of experimenters: clinicians often without technical or programming background, and researchers whose focus is more targeted on exploring new features. By helping experimenters manage reproducible pipelines, and by reducing the time and effort necessary to select adequate features for classification, \textit{HappyFeat} fills a gap in the BCI world between offline analysis of neurophysiological phenomena using pre-recorded signals, and live BCI applications. 

\textit{HappyFeat} was designed as an open-source project and its usage can be acknowledged by citing this article. \textit{HappyFeat} is continuously updated with new features and regularly maintained. Suggestions of improvements, as well as further developments, can be addressed to the corresponding authors of this article.

\section*{Acknowledgements \& Funding}
The research leading to these results has received funding from the program "Investissements d’avenir" ANR-10-IAIHU-06. The authors acknowledge support from European Research Council (ERC) under the European Union’s Horizon 2020 research and innovation program (grant agreement No. 864729).
The content is solely the responsibility of the authors and does not necessarily represent the official views of any of the funding agencies.

\clearpage

\bibliographystyle{elsarticle-num} 
\bibliography{HappyFeat_SI}

\newpage

\section*{Figures}

\clearpage

\section*{Annex}
\label{sec:annex}

\subsection*{\underline{Installation \& Requirements}}

\textit{HappyFeat} is available as a PyPI Python package: \texttt{python -m pip install happyfeat}. Alternatively, the source code can be downloaded at \url{https://github.com/Inria-NERV/happyFeat}. It requires Python 3.9 and OpenViBE 3.5.0, which can be downloaded at \url{https://openvibe.inria.fr/}. Further information can be found in the official documentation (\url{https://happyfeat.readthedocs.io/}) and in the \textit{README.md} file at the top level of the repository.

\subsection*{\underline{Classification metrics}}

Motor Imagery BCI is based on Event-Related Desynchronization/Synchronization (ERD/ERS)\cite{pfurtscheller_event-related_1999}, which corresponds to a decrease/increase of signal power between different motor-related tasks in specific frequency bands, for EEG signals in the sensorimotor cortex. This phenomenon is observable in both motor execution and motor imagery.

The discrimination between mental states, and therefore between MI tasks, is done by using a classification algorithm, with adequately selected classification features. Here, we will describe two types of features: the first is the signal's \textbf{Power Spectral Density (PSD)}, which is the most commonly used feature \cite{hwang_eeg-based_2013}; the second is Coherence-based \textbf{Functional Connectivity}\cite{cattai_characterization_2018}.

\subsection*{\underline{Power Spectral Density}}

The PSD of the recorded EEG signals is widely used as a discriminant feature in MI BCIs. One of the most common ways of estimating the PSD is by using Welch's method\cite{diez_comparative_2008}, which consists in splitting the signal into overlapping segments, applying a window function on the segments, computing the periodograms of the windowed segments (via a Discrete Fourier Transform), and finally averaging the squared magnitude of the individual periodograms.

As described in Diez et al.\cite{diez_comparative_2008}, Burg's method is more relevant in the context of EEG signals. Notably, it allows high frequency resolution for short data inputs, which makes more sense in the context of mental tasks lasting a few seconds. It implies fitting an autoregressive (AR) model to the signal, by minimizing the forward and backward prediction errors, then computing the DFT of the AR coefficients.

\subsection*{\underline{Functional Connectivity}}

As an alternative to PSD, metrics based on Functional Connectivity (FC) have been studied in the last years. FC allows the observation of brain activity as a network of time-varying connections between areas. Algorithms and analysis from network and graph theory can be applied on the connectivity matrix.

In order to estimate the degree of interaction between electrodes, the simplest measure is the \textit{coherence}, which can be seen as a generalization of correlation in the frequency domain.

With $S_{xy}(f)$ is the cross-spectrum of complex signals $x$ and $y$ at frequency $f$, and $S_{xx}(f)$ and $S_{yy}(f)$ the spectra of $x$ and $y$ at frequency $f$, we define coherence and its variants\cite{nolte_identifying_2004, cattai_characterization_2018, cattai_phaseamplitude_2021} as:

\begin{itemize}
	\item Spectral Coherence: 
	
	$COH_{xy}(f) = \frac{\left | S_{xy}(f) \right |}{\sqrt{S_{xx}(f).S_{yy}(f)} }$
	
	\item Imaginary part of Coherence: 
	
	$iCOH_{xy}(f) = \frac{\left | Im(S_{xy}(f)) \right |}{\sqrt{S_{xx}(f).S_{yy}(f)} }$
	
\end{itemize}

Computing one of those metrics yields a connectivity (or adjacency) matrix, with \textit{(nb.channels $\times$ nb.channels $\times$ nb.frequencies)} coefficients, which can also be seen as a weighted network. The information contained in this matrix can be exploited or re-formatted in many ways: for example summing all the weights associated to one edge (or channel) yields the \textit{node strength}. Other metrics can be explored, such as for example laterality, centrality, or betweenness\cite{de_vico_fallani_graph_2014}\cite{gonzalez-astudillo_network-based_2021}.

\subsection*{\underline{Processing speed}}

One way of evaluating how \textit{HappyFeat} can help in a BCI experiment is by measuring the processing times of each segment. 

The \textit{feature extraction} step is the most time consuming, and its processing time depends on the metric chosen, as computing Connectivity matrices using Autoregressive (AR) models\cite{schlogl_analyzing_2006} is more costly than PSDs, but also on the used parameters: the higher the AR model order, or the lower the time between two connectivity measurements, the costlier it gets. Of course, the number of channels has an important impact on processing times, linear for PSD, and quadratic for Connectivity.

Table \ref{tab:table_proctimes} shows a summary of the measured times, using a computer with 3 GHz 4 core CPU (8 threads) and 32 GB of RAM. Signal and extraction parameters can be found in the same table. PSD estimation is performed with Burg's method\cite{diez_comparative_2008}, Node strength is computed from Connectivity matrices using Coherence (NS-COH)\cite{cattai_characterization_2018}.

\begin{table}[h!]
	\begin{center}
		\begin{tabular}{|ccccc|}
			\hline
			\multicolumn{1}{|c|}{}							& \multicolumn{2}{c|}{\textbf{Param 1}} & \multicolumn{2}{c|}{\textbf{Param 2}} \\
			\hline
			\multicolumn{1}{|c|}{Full signal length} 		& \multicolumn{4}{c|}{7min49 (469s)}                      	\\
			\multicolumn{1}{|c|}{Number of channels}		& \multicolumn{4}{c|}{64}            			            \\
			\multicolumn{1}{|c|}{Number of trials per class}& \multicolumn{4}{c|}{20}			                        \\
			\multicolumn{1}{|c|}{Trial duration} 			& \multicolumn{4}{c|}{3s}			                        \\
			\multicolumn{1}{|c|}{Sampling frequency} 		& \multicolumn{4}{c|}{500Hz}                     		   	\\
			\multicolumn{1}{|c|}{AR model order}			& \multicolumn{4}{c|}{19}            			            \\
			\multicolumn{1}{|c|}{FFT Size}					& \multicolumn{4}{c|}{512}			                        \\
			\multicolumn{1}{|c|}{Estimation window length}	& \multicolumn{2}{c|}{0.25s} &  \multicolumn{2}{c|}{0.5s} 	\\
			\multicolumn{1}{|c|}{Overlap btw. windows}		& \multicolumn{2}{c|}{36\%} & \multicolumn{2}{c|}{20\%} 		\\
			\hline
			\multicolumn{5}{|c|}{\textbf{Processing times (s)}} \\
			\hline
			 \multicolumn{1}{|c|}{} & \multicolumn{1}{c|}{PSD}	& \multicolumn{1}{c|}{NS-COH}	& \multicolumn{1}{c|}{PSD} & \multicolumn{1}{c|}{NS-COH} \\
			\hline
			\multicolumn{1}{|c|}{Feature extraction} 		&  \multicolumn{1}{c|}{13.50} & \multicolumn{1}{c|}{50.50} & \multicolumn{1}{c|}{8.05} & \multicolumn{1}{c|}{15.55}    \\
			\multicolumn{1}{|c|}{Load data for visualization}	&  \multicolumn{1}{c|}{7.70}     & \multicolumn{1}{c|}{10.60}    & \multicolumn{1}{c|}{3.56}      & \multicolumn{1}{c|}{4.05}     \\
			\multicolumn{1}{|c|}{Classifier Training}			&  \multicolumn{1}{c|}{4.30} 	& \multicolumn{1}{c|}{7.80}     & \multicolumn{1}{c|}{4.25}      & \multicolumn{1}{c|}{7.56} \\
			\hline    
		\end{tabular}	

	\caption{\textbf{HappyFeat's main operations processing times}. }
	\label{tab:table_proctimes}	
	\end{center}
\end{table}

\clearpage

\end{document}